# Measuring the match between evaluators and evaluees: Cognitive distances between panel members and research groups at the journal level

A.I.M. Jakaria Rahman [a], Raf Guns [a], Loet Leydesdorff [b], and Tim C.E. Engels [a, c]

[a] Centre for R&D Monitoring (ECOOM), Faculty of Social Sciences, University of Antwerp, Middelheimlaan 1, B-2020 Antwerp, Belgium
[b] Amsterdam School of Communication Research (ASCoR), University of Amsterdam, PO Box 15793, 1001 NG Amsterdam, The Netherlands
[c] Antwerp Maritime Academy, Noordkasteel Oost 6, B-2030 Antwerp, Belgium

**Abstract**

When research groups are evaluated by an expert panel, it is an open question how one can determine the match between panel and research groups. In this paper, we outline two quantitative approaches that determine the cognitive distance between evaluators and evaluees, based on the journals they have published in. We use example data from four research evaluations carried out between 2009 and 2014 at the University of Antwerp.

While the barycenter approach is based on a journal map, the similarity-adapted publication vector (SAPV) approach is based on the full journal similarity matrix. Both approaches determine an entity's profile based on the journals in which it has published. Subsequently, we determine the Euclidean distance between the barycenter or SAPV profiles of two entities as an indicator of the cognitive distance between them. Using a bootstrapping approach, we determine confidence intervals for these distances. As such, the present article constitutes a refinement of a previous proposal that operates on the level of Web of Science subject categories.

# 1 Introduction

Research evaluation exercises are carried out in a number of countries across the world including the UK, Norway, Finland, Sweden, Denmark, the Netherlands, Belgium, Italy, Australia, New Zealand, Romania, China (Hong Kong), Germany, Czech Republic (Barker, 2007; Molas-Gallart, 2012; Simon & Knie, 2013; McKenna, 2015; Milat, Bauman, & Redman, 2015). The principal objective of such evaluations is to improve the quality of scientific research groups or departments within a national or regional context (Engels, Goos, Dexters, & Spruyt, 2013). In academia, publications are considered key indicators of expertise (Rybak, Balog, & Nørvåg, 2014) that help to identify the qualified or similar experts to assign papers for review (Neshati, Beigy, & Hiemstra, 2012), and to form an expert panel (Hashemi, Neshati, & Beigy, 2013).

When peer review is carried out by one or more individuals, we refer to it as individual evaluation. Although multiple individuals may evaluate the same thing, they carry out their peer review as individuals and without communication with the other evaluators. This kind of peer review is most commonly used for publications. On the other hand, panel evaluation (Coryn & Scriven, 2008; Abramo & D'Angelo, 2011) refers to a panel of experts working together in their evaluation of, e.g., a research group, an institution or a research grant application (ESF, 2011; Boyack, Chen, & Chacko, 2014). Contrary to individual evaluation, this kind of peer review presupposes frequent contact and communication between the evaluators. It may include site visits by the expert panel members for data gathering and evaluations (Borum & Hansen, 2000; Hansson, 2010; Lawrenz, Thao, & Johnson, 2012). Mixed forms of both types occur frequently. In general, however, the current paper is especially concerned with the peer review process in the context of expert panel evaluation of research groups.

A downside of the peer review process can be the absence of an adequate methodology to find relevant experts (Hofmann, Balog, Bogers, & de Rijke, 2010; Gould, 2013; Lee, Sugimoto, Zhang, & Cronin, 2013; Berendsen, de Rijke, Balog, Bogers, & Bosch, 2013; Oleinik, 2014; Buckley, Sciligo, Adair, Case, & Monks, 2014). The peer review process is an established component of professional practice, and often the expert is anonymous to the unit of assessment. Expert panel review is a standard practice for evaluating research groups (Nedeva, Georghiou, Loveridge, & Cameron, 1996; Rons, De Bruyn, & Cornelis, 2008; Butler & McAllister, 2011; Lawrenz et al., 2012; Milat et al., 2015), and for research proposals submitted to research funding organizations (Wessely, 1998; van den Besselaar & Leydesdorff, 2009; Li & Agha, 2015; Wang & Sandström, 2015; Pina, Hren, & Marušić, 2015). In expert panel evaluation, however, the panel members are visible, and hence the units of assessment themselves can judge the expertise of the panel member and the expert panel in relation to their research domain.

The exponential growth of research literature indicates the growth of specialized disciplines (Sobkowicz, 2015) as well as the growth of databases themselves. Therefore, an individual panel member may have sufficient expertise in a given field, but collaborative evaluation together with peers is crucial unless and until the individual panel member covers the expertise of the research groups. In expert panel evaluation the entire panel preferably has expertise on the discipline of the research groups; otherwise the trustworthiness of the evaluation is open for discussion (Engels et al., 2013). In our opinion, a methodology is required to set the standard for most appropriate expert panel composition. One of the main factors that need to be taken into account is the cognitive distance between an expert panel and research groups (Rahman, Guns, Rousseau, & Engels, 2014, 2015; Wang & Sandström, 2015).

The concept of cognitive distance has been developed in the academic literature by Nooteboom and colleagues (Nooteboom, 1999, 2000; Nooteboom, Van Haverbeke, Duysters, Gilsing, & van den Oord, 2007). Cohen & Levinthal (1989, 1990) explained the process by which an individual or organization, by extrapolation can integrate and reuse knowledge from outside sources in research and development, while Nooteboom uses these ideas to define the concept of cognitive distance between individuals and organizations. Nooteboom (2000, p. 73) defines cognitive distance as "a difference in cognitive function. This can be a difference in domain, range, or mapping. People could have a shared domain but a difference of mapping: two people can make sense of the same phenomena, but do so differently". Thus, cognitive distance describes how two individuals – and, by extension, organizations or groups of individuals – are different, in terms of knowledge, but also in the way they perceive and interpret external phenomena. In this paper, and like Wang & Sandström (2015), we consider the publication profile of the involved researchers to determine cognitive distance between people and groups of people. For example, if a panel member and a research group have a publication in the same or similar journals it indicates a smaller cognitive distance between them. Hence, we measured cognitive distance between panel members and research groups based on how often they have published in the same or similar journals.

In this paper, we study the problem of composing an expert panel, such that the individual panel members' expertise covers the specific subdomains in the discipline where the units of assessment have publications. Since 2007, the University of Antwerp (Belgium) expert panel evaluation has included site visits by the expert panel members. One expert panel is accountable for a specific department, e.g. Biology, and evaluates all the research groups belonging to this department. Following the Dutch Standard Evaluation Protocol (SEP: VSNU, 2003, 2009), the panels assessed the quality, the productivity, the relevance, and the viability of the research groups without, however, a direct influence on the resource (re)allocation to those groups. The panel members are recognized independent international specialists in at least one of the fields addressed by the department under evaluation, and have no prior joint affiliations, no co-publications, no common projects etc. with the assessed research groups. The research groups consist of professors (of all ranks), research and teaching assistants, and researchers (PhD students and postdocs). These evaluations consider the entire research groups scientific activity for a specific period, typically eight years. We previously explored expertise overlap between panel and research groups through publishing in the same or similar WoS subject categories (Rahman et al., 2014, 2015; Ronald Rousseau, Rahman, Guns, & Engels, 2016) . Since one subject category may comprise a wide array of different subfields and topics (Bornmann, Mutz, Marx, Schier, & Daniel, 2011), it is up for discussion how relevant it is to have panel members and research group members publishing in the same subject categories. As journals cover more closely related subfields and topics (Tseng & Tsay, 2013), we present a journal level analysis to explore the issue.

The analysis relies on the journal similarity matrix and the overlay map derived from it. Science overlay maps (Rafols, Porter, & Leydesdorff, 2010) have received considerable attention from the field of informetrics (Grauwin & Jensen, 2011; Boyack & Klavans, 2014; Fields, 2015; Chen, Arsenault, Gingras, & Lariviere, 2015; Gorjiara & Baldock, 2014). We present two bibliometric approaches to assess the cognitive distances between research groups in the Department of Biomedical Sciences, Veterinary Sciences, Pharmaceutical Sciences, Biology, and the respective expert panels based on research evaluations carried out at the University of Antwerp. We have used the data collected in the frame of research evaluation by the University of Antwerp. We explore the cognitive distance between expert panel and research groups. The research questions are:

1) How can one quantify the cognitive distances between two entities using the journals in which they have published? How can one estimate the uncertainty inherent to these cognitive distances?
2) To what extent was each individual research group's expertise covered by the panel's expertise?
3) To what extent does each individual panel member's expertise cover the individual research groups?

## 2 Data

In this paper, we consider data from the research assessments of all the research groups belonging to four departments of the University of Antwerp, Belgium. These are the 2014 assessment of 15 research groups belonging to the department of Biomedical Sciences, the 2014 assessment of the three research groups of the Veterinary Sciences department, the 2009 assessment of the 10 research groups of the department of Pharmaceutical Sciences, and the 2011 assessment of the nine research groups of the department of Biology. The group names will be standardized using the first four letters of the corresponding department, for example BIOM-A for Biomedical Sciences group A, VETE-C for Veterinary Sciences group C, etc. The reference period encompasses eight years preceding the evaluation. We considered all the articles, letters, notes, proceedings papers, and reviews by the research groups published during the reference period and included in the Science Citation Index Expanded (SCIE), and the Social Sciences Citation Index (SSCI) of the WoS in the evaluation.

**Table 1 Publication profile of the research groups.**

| Group code | Number of Journals | Number of Publications | Group code | Number of Journals | Number of Publications |
|---|---|---|---|---|---|
| **Biomedical Sciences (2006-2013)** | | | **Pharmaceutical Sciences (2001-2008)** | | |
| BIOM-A | 55 | 96 | PHAR-A | 22 | 40 |
| BIOM-B | 27 | 43 | PHAR-B | 32 | 62 |
| BIOM-C | 47 | 107 | PHAR-C | 35 | 61 |
| BIOM-D | 95 | 201 | PHAR-D | 17 | 32 |
| BIOM-E | 34 | 70 | PHAR-E | 42 | 64 |
| BIOM-F | 17 | 27 | PHAR-F | 21 | 34 |
| BIOM-G | 115 | 241 | PHAR-G | 31 | 67 |
| BIOM-H | 29 | 50 | PHAR-H | 27 | 39 |
| BIOM-I | 55 | 89 | PHAR-I | 10 | 29 |
| BIOM-J | 27 | 47 | PHAR-J | 9 | 11 |
| BIOM-K | 43 | 74 | All groups together | **180** | **376** |
| BIOM-L | 11 | 12 | | | |
| BIOM-M | 67 | 164 | | | |
| BIOM-N | 43 | 114 | | | |
| BIOM-O | 32 | 60 | | | |
| All groups together | **476** | **1,213** | | | |
| **Veterinary Sciences (2006-2013)** | | | **Biology (2004-2010)** | | |
| VETE-A | 102 | 144 | BIOL-A | 53 | 168 |
| VETE-B | 33 | 41 | BIOL-B | 33 | 58 |
| VETE-C | 21 | 52 | BIOL-C | 75 | 212 |
| All groups together | **146** | **231** | BIOL-D | 68 | 176 |
| | | | BIOL-E | 69 | 169 |
| | | | BIOL-F | 35 | 58 |
| | | | BIOL-G | 139 | 280 |
| | | | BIOL-H | 42 | 67 |
| | | | BIOL-I | 52 | 86 |
| | | | All groups together | **372** | **1,156** |

Table 1 lists the number of publications of the research groups. The numbers reported for all groups together are smaller than the sum of the individual research groups' publication or journal counts, because of joint publications between groups.

**Table 2 Publication profile of the panel members.**

| Panel member code | Number of journals | Number of publications | Panel member code | Number of journals | Number of publications |
|---|---|---|---|---|---|
| **Biomedical Sciences** | | | **Pharmaceutical Sciences** | | |
| BIOM-PM1 | 78 | 153 | PHAR-PM1 | 39 | 122 |
| BIOM-PM2 | 81 | 201 | PHAR-PM2 | 93 | 351 |
| BIOM-PM3 | 79 | 261 | PHAR-PM3 | 91 | 259 |
| BIOM-PM4 | 86 | 240 | PHAR-PM4 | 67 | 124 |
| BIOM-PM5 | 37 | 74 | PHAR-PM5 | 86 | 180 |
| BIOM-PM6 | 35 | 109 | All Panel members together | **300** | **1,032** |
| BIOM-PM7 | 68 | 194 | | | |
| BIOM-PM8 | 32 | 101 | | | |
| All Panel members together | **395** | **1,319** | | | |

| Panel member code | Number of journals | Number of publications | Panel member code | Number of journals | Number of publications |
|---|---|---|---|---|---|
| **Veterinary Sciences** | | | **Biology** | | |
| VETE-PM1 | 50 | 313 | BIOL-PM1 | 48 | 146 |
| VETE-PM2 | 66 | 121 | BIOL-PM2 | 49 | 177 |
| VETE-PM3 | 46 | 272 | BIOL-PM3 | 35 | 76 |
| VETE-PM4 | 53 | 131 | BIOL-PM4 | 49 | 185 |
|  |  |  | BIOL-PM5 | 76 | 262 |
| All Panel members together | **200** | **837** | All Panel members together | **217** | **792** |

Table 2 lists the number of publications of the panel members. The entire WoS publication record of the individual panel members up to the year of assessment was taken into account. The Veterinary Sciences and Biomedical Sciences panels were composed of four and eight members respectively. Both the Pharmaceutical Sciences and Biology panels were composed of five members including the chair. There are no co-authored publications between panel members of Veterinary Sciences. Pharmaceutical Sciences, Biomedical Sciences, and Biology panel members have 4, 14, and 54 publications in collaboration between two or more panel members respectively.

## 3 Methods

### 3.1 Journal similarity matrix and maps

Our method is based on the assumption that the cognitive distance between entities decreases as they have more publications in the same or similar journals, since journals cover closely related subfields and topics. The similarity between journals should be taken into account: if a panel member publishes in different journals than the research groups, they may still have relevant expertise, if their publications are in similar or closely related journals. This requirement rules out a number of approaches, including direct comparison of the top n journals in which two entities have published and correlations between journal portfolios.

We have harvested data from Thomson Reuters' WoS Journal Citation Reports (JCR) of the Science and Social Science Editions 2011. An aggregated journal-journal citation matrix of 10,675 journals[1] was constructed with a grand total of 35,295,459 citations over the entire matrix, which was subsequently normalized in the citing direction. The distances between journals are calculated using the cosine similarity between their citing distributions respectively (see Leydesdorff, Rafols, & Chen (2013) for details). The resulting journal similarity matrix can be considered as an adjacency matrix, and thus is equivalent to a weighted network where similar journals are linked and link weights increase with similarity strength. At the moment, it is not yet entirely clear how intense citation traffic around journals such as PLoS ONE (Leydesdorff & de Nooy, 2015) affects the journal similarity matrix.

---

[1] The Science and Social Science Editions 2011 contain 8,281 and 2,943 journals respectively. Of these journals, 549 are contained in both databases.

The journal similarity matrix consists of $10,675^2 = 113,955,625$ cells. The matrix was stored using the HDF5 format (Hierarchical Data Format version 5), which was found to be the most efficient way of storing the data in terms of speed and memory requirements.

We used the full title of the journals for matching journals in the panel's publication list with journals in the research groups' publication lists. However, journals are not static entities and may undergo a name or organizational changes over time. Possible changes include:

- The journal title is changed, shortened or extended;
- Two or more journals merge into a new journal;
- One journal splits into two or more new journals;
- A journal is excluded from the WoS, discontinued, or not listed during the construction of the aggregated journal-journal citation matrix.

While cross-matching, we found 165 journals in our data set that belong to any of the above mentioned categories. We developed the following guidelines to handle these uniformly:

- If journal A is renamed to B then treat both as equivalent.

- If journals A1 and A2 are merged into journal B, we treat both A1 and A2 as equivalent to B.

- If journal X splits into multiple journals, we look up which research groups or panel members have publications in journal X and determine which of the new journals best corresponds to the specialty of the authors, then change all occurrences of the journals in the WoS exported data with the best fitting latter journals. This was necessary in 15 cases; each time the decision was quite clear.

- If a journal is discontinued or excluded from WoS, or not included in the aggregated journal-journal citation matrix and there is no equivalent for some other reason, then it is removed from the sample.

From the journal similarity matrix, one can construct a global journal map (Leydesdorff & Rafols, 2012), in which similar journals are located more closely together. When used as a portfolio map, the size of the nodes depends on the number of publications in each node, and helps to compare the degree of overlap of multiple entities visually (Leydesdorff, Heimeriks, & Rotolo, 2015). The overlay of research group and panel publications can be visualized on the global journal map based on the retrieved publications data, using the visualization program VOSviewer (van Eck & Waltman, 2010). However, in the process of visualization, the multi-dimensional space is reduced to a projection in two dimensions. Moreover, comparison of overlay maps is difficult, specifically when the journals are located (very) closely to one another or when a panel member or research group has published in many different journals. Therefore, we will explore two approaches to create a 'profile' of a panel member or research group: (i) barycenters on the overlay map (Rahman et al., 2015), and (ii) similarity-adapted publication vectors or SAPVs (Rousseau et al., 2016). Subsequently, we can determine and compare the distances between entities, with overlay maps providing additional qualitative context.

*3.2   Barycenter and distance calculation*

Our barycenter approach is based on the journal map. The barycenter is an entity's weighted average location on the map. More specifically, an entity's barycenter is the center of weight (Rousseau, 1989, 2008) of the journals in which it has published, where a journal's weight is the entity's number of publications in that journal. The barycenter is defined as the point $C = (C_1, C_2)$, where

$$C_1 = \frac{\sum_{j=1}^{N} m_j L_{j,1}}{T} \; ; \; C_2 = \frac{\sum_{j=1}^{N} m_j L_{j,2}}{T} \qquad (1)$$

Here, $L_{j,1}$ and $L_{j,2}$ are the horizontal and vertical coordinates of journal *j* on the map, $m_j$ is the number of publications in journal *j*, and $T = \sum_{j=1}^{N} m_j$ is the total number of publications of the entity. For further elaboration on the barycenter, we refer to (Rousseau, 1989; Jin & Rousseau, 2001; Verleysen & Engels, 2013, 2014).

The Euclidean distance between points $C = (C_1, C_2)$ and $D = (D_1, D_2)$ is calculated as follows:

$$d = \sqrt{(C_1 - D_1)^2 + (C_2 - D_2)^2}. \qquad (2)$$

Many different algorithms and layout techniques have been developed for visualization of matrices. Rahman et al., (2015) found that at least two strongly different techniques – Kamada-Kawai (Kamada & Kawai, 1989) and VOS (van Eck & Waltman, 2007; van Eck, Waltman, Dekker, & van den Berg, 2010) – yielded very similar results in terms of barycenter distances. The journal map used in this paper was created using the VOS algorithm as implemented in VOSviewer (van Eck & Waltman, 2010). Subsequently, we determine and compare the cognitive distance between entities, with overlay maps providing additional qualitative context through visual comparison. In the Results section, we present several overlay maps (see figure 1 to 4) including barycenters and corresponding confidence regions (see section 3.4 for details). These maps are zoomed in to better highlight places of interest, hence independent of the zoom level of the figures.

*3.3   Similarity-adapted publication vectors (SAPV) and distance calculation*

In earlier work (Rahman et al., 2015) we introduced a technique we referred to as 'N-dimensional barycenters'. This terminology, as well as the normalization used, was corrected by Rousseau et al., (2016) who introduced the idea of similarity-adapted publication vectors (SAPVs). Whereas a regular publication vector simply contains publication counts per journal (or subject category), in a similarity-adapted publication vector these counts are adapted to account for similarity between journals. We will use normalized SAPVs, such that there is scale invariance and publication vectors of entities of varying size can be meaningfully compared.

We calculate SAPVs for each entity, starting from the original journal similarity matrix, where $N = 10{,}675$ is the number of rows or columns in the matrix. Based on their respective SAPVs, the distance can be calculated between the expert panel, panel members, groups, and separate groups.

A similarity-adapted publication vector is determined as the vector $C = (C_1, C_2, \ldots, C_N)$, where:

$$C_k = \frac{\sum_{j=1}^{N} m_j s_{jk}}{\sum_{i=1}^{N} \sum_{j=1}^{N} m_i s_{i,j}} \quad (3)$$

Here $s_{j,k}$ denotes the $k$-th coordinate of journal $j$ and $m_j$ is the number of publications in journal $j$. The numerator of Equation (3) is equal to the $k$-th element of $S * M$, the multiplication of the similarity matrix $S$ and the column matrix of publications $M = (m_j)_j$. The denominator is the $L_1$-norm of the unnormalized vector.

Subsequently, we determine the distance between the expert panel as a whole and individual panel members on the one hand, and the department (the combined groups), and individual groups on the other. The Euclidean distance between vectors a and b in $\mathbf{R}^N$ is:

$$d(a,b) = \sqrt{(a_1 - b_1)^2 + \cdots + (a_N - b_N)^2} \quad (4)$$

Although the matrix and vectors are large, the calculation of SAPV and distances is relatively fast, due to the use of efficient matrix procedures implemented in NumPy and SciPy.[2]

Both the SAPV approach and barycenter approach can be used to determine an entity's 'profile'. One can then calculate the distance between profiles as an indicator of cognitive distance. For each research group we find the shortest distance to one of the panel members. We use the average and standard deviation of the shortest distances as a comparative measure. All the distances are shown up to the third decimal. The distances are arbitrary units on a ratio scale (Egghe & Rousseau, 1990). Hence, one can meaningfully compare them in terms like 'x is twice as large as y'.

*3.4   Confidence intervals*

The barycenter and SAPV approaches determine cognitive distance on the basis of the journals in which the groups and panel members have published. However, such information is not entirely deterministic; it is, for instance, dependent on the database used as well as environmental factors like the speed with which a journal processes a submission. It logically follows that small differences in Euclidean distances bear little meaning. To study this problem in a more systematic way, we employ a bootstrapping approach in order to determine 95% confidence intervals (CIs) to each Euclidean distance (both between barycenters and SAPVs). If two CIs do not overlap, the difference between the distances is statistically significant at the 0.05 level. Although it is possible for overlapping CIs to have a statistically significant difference between the corresponding distances, the difference between the distances is less likely to have practical meaning.

---
[2] http://www.numpy.org/ and http://scipy.org

Bootstrapping (Efron & Tibshirani, 1998) is a simulation-based method for estimating standard error and confidence intervals. Bootstrapping depends on the notion of a *bootstrap sample*. To determine a bootstrap sample for a panel member or research group with N publications, we randomly sample with replacement N publications from its set of publications. In other words, the same publication can be chosen multiple times. Some publications in the original data set will not occur in the bootstrap data set, whereas others will occur once, twice or even more times. From the bootstrap sample, one can calculate a *bootstrap replication*, in our case a barycenter using formula (1) or SAPV using formula (3).

By generating a large amount of independent bootstrap samples (in our case 1000) and each time calculating the bootstrap replication, we can approximate the variability within the data set. Since we have a two-sample problem (distance between two entities; Efron & Tibshirani, 1998, Ch. 8), we calculate the distances between pairs of bootstrap replications, from which we obtain a CI using a bootstrap percentile approach (Efron & Tibshirani, 1998, Ch. 13). A more detailed explanation and implementation of our method is available on Github (http://nbviewer.jupyter.org/gist/rafguns/6fa3460677741e356538337003692389).

The bootstrap replications of barycenters are also used to add a 95% confidence region for each barycenter to the maps. For each barycenter we have a cloud of 1000 points (bootstrapped barycenters) surrounding it. The confidence region is an ellipse that covers 95% of the bootstrapped barycenters and is obtained using an implementation by Kington (2014). The larger the confidence region, the less stable the barycenter is. Although the CI of the distance between two barycenters and their confidence regions are related, the two should not be conflated. In particular, we stress that overlapping confidence regions as seen in e.g. Fig. 1 does not correspond to overlap between CIs for distances.

## 4  Results

We present the results in four parts. In the first (section 4.1) and the second part (section 4.2), we will discuss the results of Euclidean distances between barycenters and distances between SAPVs respectively. In the third part (section 4.3), we discuss the confidence intervals of both the approaches. However, for the intelligibility we show all the relevant tables of the Euclidean distance of barycenter and SAPV in the section 4.1 and 4.2, where the confidence intervals are included through the typography of the values. In the last part (section 4.4), we make a comparison between our two approaches.

### *4.1  Barycenter and distances*

For each discipline, the barycenters of the panel, panel members, individual research groups and department, as well as Euclidean distances between barycenters are calculated. For each research group we also calculate the average shortest distance to one of the panel members. The visualizations of barycenters and their confidence regions are added to the overlay maps.

*Biomedical Sciences*

Table 3 provides data on the distances between the barycenters of the panel and its members on the one hand and those of the department and individual research groups on the other. The Biomedical panel is very near to BIOM-F (0.064), while BIOM-G (0.396), BIOM-H (0.354), BIOM-L (0.383), and BIOM-N (0.371) are almost 5 to 6 times farther away from the panel than BIOM-F. BIOM-C (0.146), BIOM-D (0.109), BIOM-I (0.133) groups are situated comparatively close to the panel's coordinates, while BIOM-E (0.263) is found at a considerable distance from the panel's barycenter.

In Table 3, the average of the shortest distance between the Biomedical Sciences groups and panel members is 0.132 (SD 0.06) and can be used as a measure of the fit between the expertise of the Biomedical Sciences panel and the research groups. Groups BIOM-G, BIOM-H, BIOM-M, and BIOM-N are situated moderately far away from the panel's coordinates, but PM2 and PM6 are located in their immediate neighborhood.

Similar conclusions can be drawn from the visualization in Figure 1. Here, 'PM' stands for 'panel member', 'Panel' represents the barycenter location of the publication profile of the entire panel, and 'Groups' does the same for the research groups taken together (the department). The advantage of the visual representation consists in providing an easily interpretable overview of how the panel and research groups relate, which is much less straightforward from a table of distances.

**Table 3 Euclidean distances between barycenters of Biomedical Sciences individual research groups, panel members, panel and groups together in the journal VOS-map.**

|       | Groups | BIOM-A | BIOM-B | BIOM-C | BIOM-D | BIOM-E | BIOM-F | BIOM-G | BIOM-H | BIOM-I | BIOM-J | BIOM-K | BIOM-L | BIOM-M | BIOM-N | BIOM-O |
|-------|--------|--------|--------|--------|--------|--------|--------|--------|--------|--------|--------|--------|--------|--------|--------|--------|
| Panel | 0.177  | 0.225  | 0.132  | 0.146  | 0.109  | 0.263  | 0.064  | 0.396  | 0.354  | 0.133  | 0.303  | 0.268  | 0.383  | 0.312  | 0.371  | 0.282  |
| PM1   | 0.265  | 0.350  | 0.180  | 0.224  | **0.110** | 0.242 | **0.081** | 0.473 | 0.319 | **0.159** | 0.445 | 0.387 | **0.471** | 0.397 | 0.436 | 0.344 |
| PM2   | 0.085  | **0.176** | **0.038** | **0.046** | 0.201 | 0.177 | **0.119** | **0.302** | 0.267 | **0.234** | 0.297 | 0.208 | **0.294** | **0.221** | **0.272** | **0.181** |
| PM3   | 0.413  | 0.390  | 0.397  | 0.397  | 0.241  | 0.530  | 0.303  | 0.611  | 0.621  | **0.194** | 0.356 | 0.438 | 0.586 | 0.527 | 0.599 | 0.522 |
| PM4   | 0.389  | 0.391  | 0.355  | 0.365  | **0.168** | 0.479 | 0.243 | 0.600 | 0.568 | **0.119** | 0.390 | 0.440 | 0.580 | 0.515 | 0.582 | 0.498 |
| PM5   | 0.149  | **0.250** | 0.058 | 0.107 | 0.183 | 0.144 | 0.095 | 0.348 | **0.233** | 0.227 | 0.371 | **0.280** | 0.348 | 0.274 | 0.311 | 0.220 |
| PM6   | 0.189  | **0.295** | 0.177 | 0.184 | 0.383 | **0.072** | 0.295 | **0.236** | **0.086** | 0.426 | 0.442 | **0.291** | 0.258 | **0.207** | **0.187** | **0.135** |
| PM7   | 0.251  | 0.367  | 0.173  | 0.217  | 0.282  | **0.103** | 0.209 | 0.395 | **0.148** | 0.331 | 0.500 | 0.385 | **0.407** | 0.342 | 0.348 | 0.271 |
| PM8   | 0.275  | **0.171** | 0.363 | 0.314 | 0.497 | 0.445 | 0.445 | **0.238** | 0.502 | 0.504 | **0.154** | **0.140** | **0.199** | 0.213 | 0.271 | 0.281 |

Shortest distances between a group and a panel member are underlined and printed in bold. Average shortest distances is 0.132 (SD 0.06). Distances whose confidence intervals overlap with that of the shortest distance are in bold.

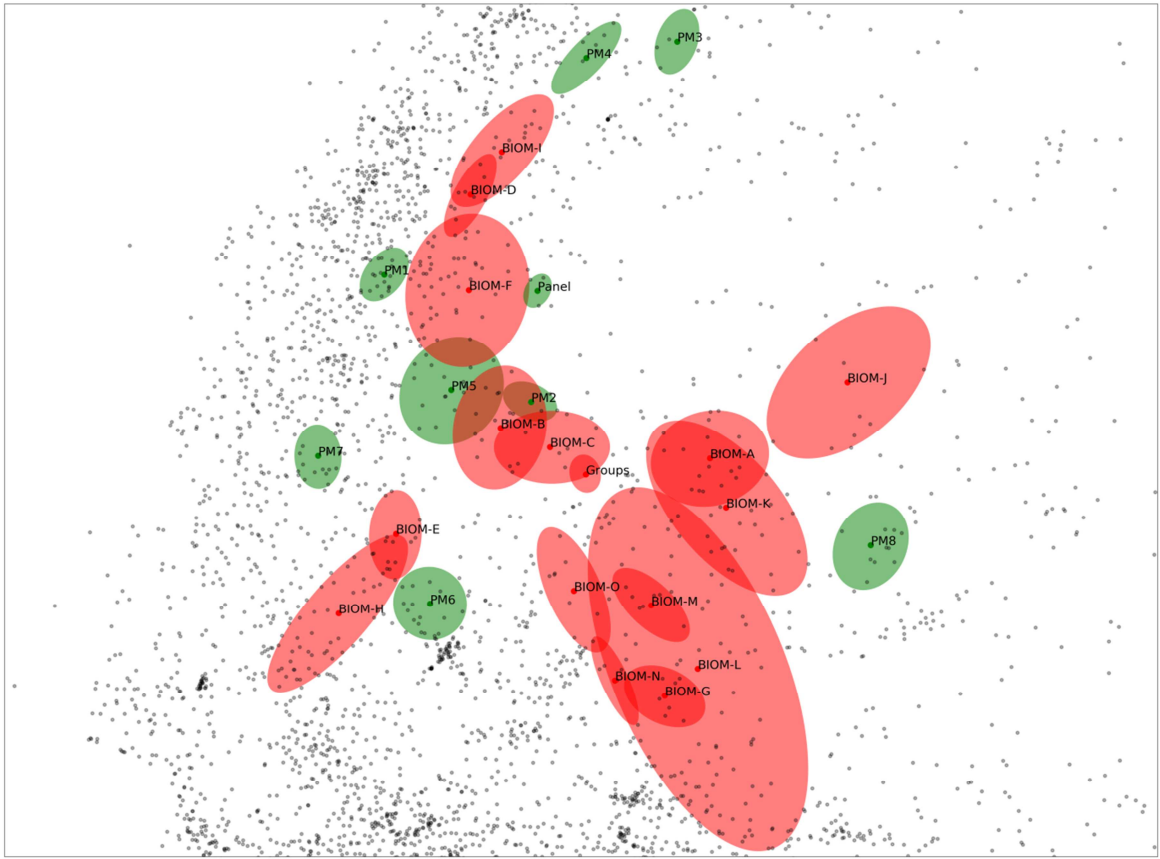

**Fig. 1 Barycenter overlay map of Biomedical Sciences panel, panel members (PM), research groups and research groups together (groups) with their confidence regions.**

*Veterinary Science*

Table 4 provides data on the distances between the Veterinary science panel's barycenter and those of the individual research groups. The Veterinary panel is the closest to VETE-B, while VETE-A is 1.9 times and VETE-C is 1.7 times farther away from the panel than VETE-B.

**Table 4 Euclidean distances between barycenters of Veterinary Sciences individual research groups, panel members, panel and groups together in the journal VOS-map.**

|       | Groups | VETE-A | VETE-B | VETE-C |
|-------|--------|--------|--------|--------|
| Panel | 0.092  | 0.179  | 0.076  | 0.156  |
| PM1   | 0.178  | 0.260  | **0.160** | **0.124** |
| PM2   | 0.088  | **0.141** | **0.108** | 0.227  |
| PM3   | 0.195  | 0.273  | **0.182** | 0.145  |
| PM4   | 0.306  | 0.272  | 0.310  | 0.469  |

Shortest distances between a group and a panel member are underlined and printed in bold. Average shortest distances is 0.124 (SD 0.013). Distances whose confidence intervals overlap with that of the shortest distance are in bold.

The overlay map (Fig. 2) shows that the panel members are generally quite close to the research groups. Only PM4 is located a bit further away from the groups. Although the fit in this case is fairly good, an even better fit could be obtained if PM4 were replaced with a different person with publications in journals that are more closely related to the groups' publication profile.

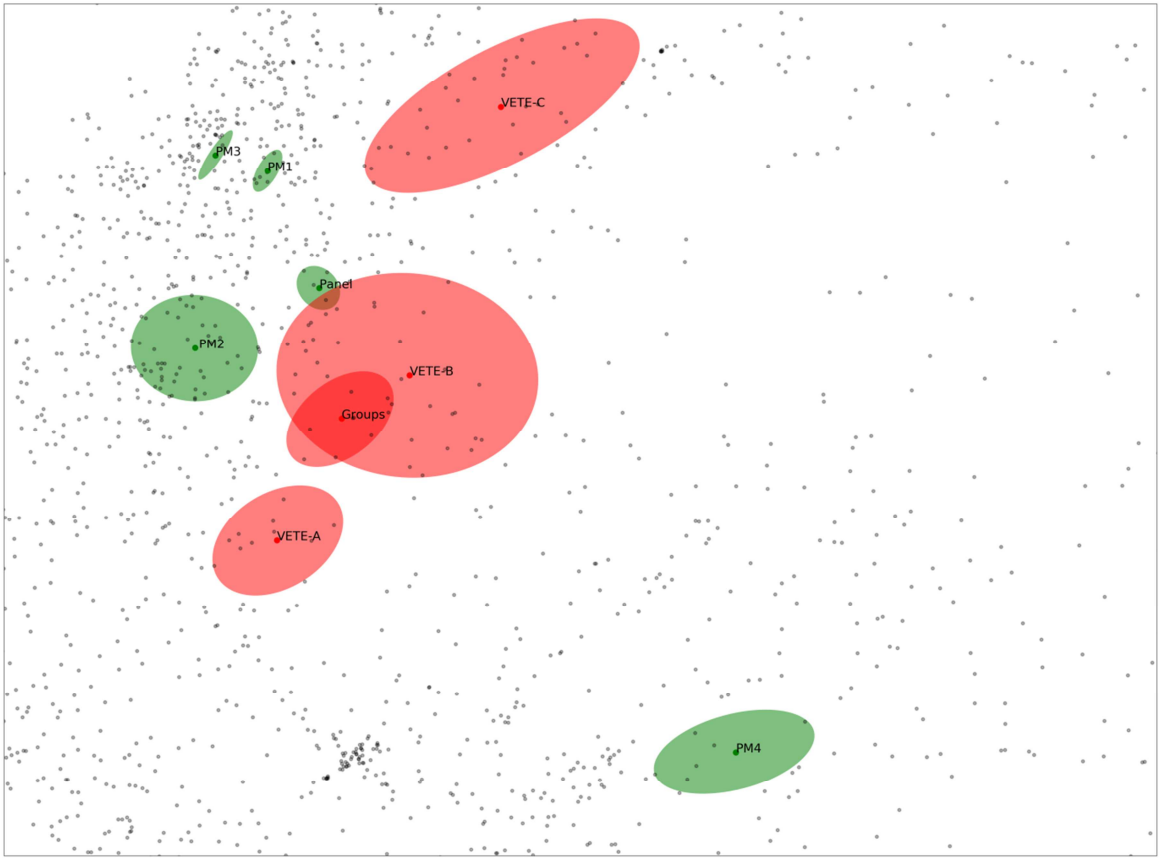

**Fig. 2** Barycenter overlay map of Veterinary Sciences panel, panel members (PM), research groups and research groups together (groups) with their confidence regions.

*Pharmaceutical Sciences*

Table 5 provides data on the distances between the Pharmaceutical sciences panel's barycenter and individual research groups. Fig. 3 visualizes the situation. The average of the shortest distance between the Pharmaceutical groups and panel members is 0.143. PHAR-C (0.536) and PHAR-I (0.769) are 5.58, and 8.01 times farther away respectively from the panel than PHAR-D (0.096). PHAR-B (0.240), PHAR-F (0.239), PHAR-H (0.120) are situated comparatively close to the panel's coordinates, while PHAR-A (0.410) and PHAR-J (0.495) are found at a considerable distance from the panel's barycenter. The case of PHAR-A reinforces our assertion that the mere overlap of journals is not sufficient to quantify the cognitive distance: although 60% of the journals in which this group has published are also covered by the panel, it is located relatively far away from the panel.

PHAR-I and the panel do not share any common journals. PHAR-I is located far away from the panel as a whole as well as from any individual panel member. In summary, the Pharmaceutical Sciences panel appears to cover most research groups adequately, with the exception of two.

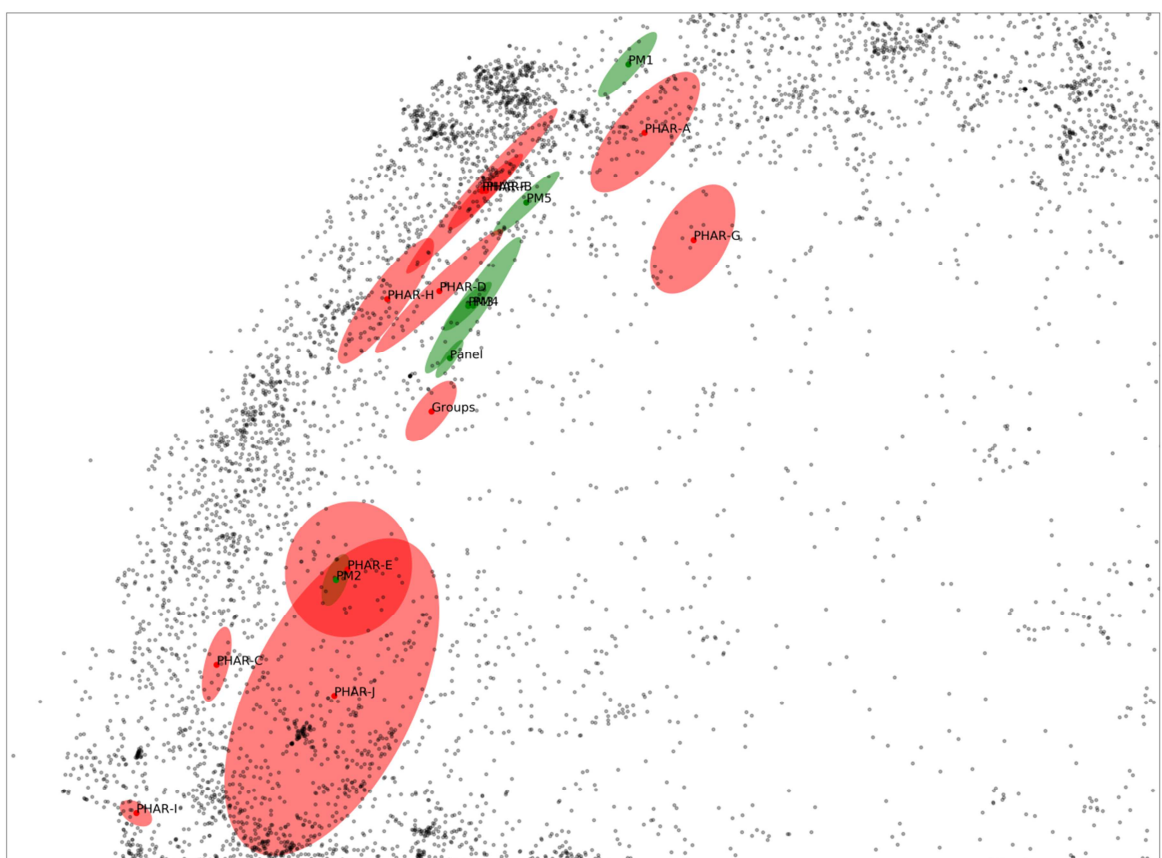

**Fig. 3 Barycenter overlay map of Pharmaceutical Sciences panel, panel members (PM), research groups and research groups together (groups) with their confidence regions.**

**Table 5 Euclidean distances between barycenters of Pharmaceutical Sciences research groups, panel members, panel and groups together in the journal VOS-map.**

|  | Groups | PHAR-A | PHAR-B | PHAR-C | PHAR-D | PHAR-E | PHAR-F | PHAR-G | PHAR-H | PHAR-I | PHAR-J |
|---|---|---|---|---|---|---|---|---|---|---|---|
| Panel | 0.078 | 0.410 | 0.240 | 0.536 | 0.096 | 0.325 | 0.239 | 0.381 | 0.120 | 0.769 | 0.495 |
| PM1 | 0.559 | **0.101** | 0.267 | 1.017 | 0.413 | 0.807 | **0.271** | **0.262** | 0.471 | 1.251 | 0.972 |
| PM2 | 0.268 | 0.750 | 0.581 | **0.205** | 0.428 | **0.021** | 0.579 | 0.689 | 0.398 | **0.429** | **0.162** |
| PM3 | 0.156 | **0.339** | **0.163** | 0.610 | **0.043** | 0.402 | **0.162** | **0.332** | **0.110** | 0.844 | **0.573** |
| PM4 | 0.160 | **0.332** | **0.161** | 0.616 | **0.052** | 0.408 | **0.160** | **0.322** | **0.120** | 0.850 | **0.577** |
| PM5 | 0.318 | **0.186** | **0.057** | 0.773 | **0.170** | 0.566 | **0.062** | **0.242** | 0.233 | 1.008 | 0.735 |

Shortest distances between a group and a panel member are underlined and printed in bold. Average shortest distances is 0.143 (SD 0.124). Distances whose confidence intervals overlap with that of the shortest distance are in bold.

*Biology*

Table 6 and Fig. 4 provide data on the distances between the Biology panel's barycenter and individual research groups. The average of the shortest distances between the Biology groups and panel members is 0.09. The Biology panel as a whole is closer to BIOL-I (0.087) and BIOL-G (0.065). BIOL-B (0.242), BIOL-C (0.271), BIOL-D (0.228) and BIOL-H (0.262) are the furthest from the panel. BIOL-A and BIOL-E are found at a considerable distance from the panel's barycenter but PM2 is in their immediate neighborhood.

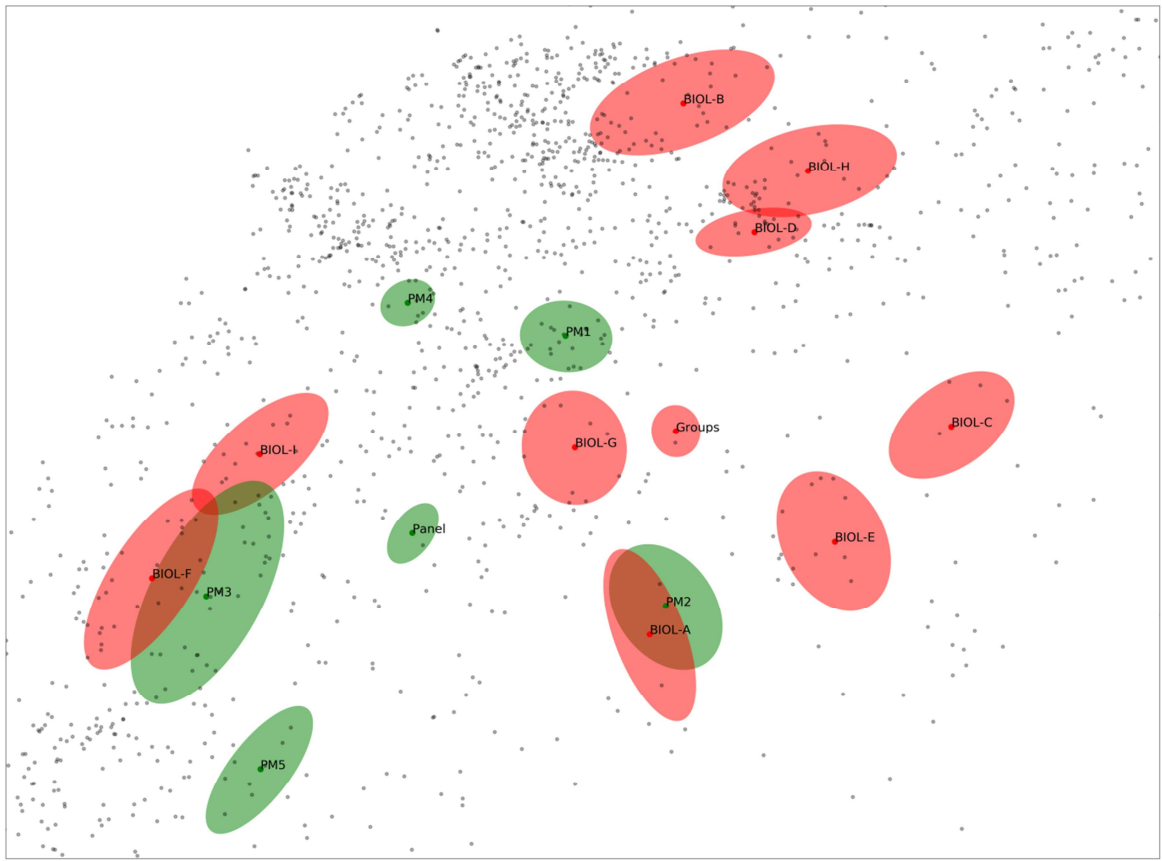

**Fig. 4 Barycenter overlay map of Biology panel, panel members (PM), research groups and research groups together (groups) with their confidence regions.**

**Table 6 Euclidean distances between barycenters of Biology individual research groups, panel members, panel and groups together in the journal VOS-map.**

|     | Groups | BIOL-A | BIOL-B | BIOL-C | BIOL D | BIOL-E | BIOL-F | BIOL-G | BIOL-H | BIOL-I |
|-----|--------|--------|--------|--------|--------|--------|--------|--------|--------|--------|
| Panel | 0.136 | 0.128 | 0.242 | 0.271 | 0.220 | 0.208 | 0.136 | 0.087 | 0.262 | 0.087 |
| PM1 | 0.072 | 0.154 | **_0.125_** | **0.198** | **_0.105_** | 0.169 | 0.239 | **_0.056_** | **_0.146_** | **0.164** |
| PM2 | 0.087 | **_0.016_** | 0.249 | **0.168** | 0.190 | **_0.090_** | 0.257 | **0.091** | **0.227** | 0.217 |
| PM3 | 0.248 | 0.223 | 0.336 | 0.382 | 0.326 | 0.316 | **_0.029_** | 0.199 | 0.368 | **_0.075_** |
| PM4 | 0.148 | 0.205 | **0.163** | 0.279 | 0.175 | 0.245 | 0.187 | **0.110** | **0.211** | **0.106** |
| PM5 | 0.253 | 0.195 | 0.374 | 0.373 | 0.348 | 0.297 | **0.104** | 0.211 | 0.390 | **0.145** |

Shortest distances between a group and a panel member are underlined and printed in bold. Average shortest distances is 0.09 (SD 0.05). Distances whose confidence intervals overlap with that of the shortest distance are in bold.

### 4.2 *Similarity-adapted publication vectors (SAPV) and distances*

For each discipline, the SAPV of the panel, panel members, individual research groups and department, as well as Euclidean distances between SAPVs are calculated. For each research group we also calculate the average shortest distance to one of the panel members.

Biomedical SciencesTable 7 provides data on the Euclidean distances between (SAPVs of) Biomedical groups, panel and panel members. BIOM-F, and BIOM-I are in the immediate neighborhood of the panel while BIOM-N (0.010) is located farest away from the panel. PM2 and PM5 are closer to nine and ten research groups respectively, while PM8 is situated moderately far away from all the research groups. The average of the shortest distances between the Biomedical Sciences groups and panel members is 0.005 (SD 0.002), which can be used as a measure of the fit between the expertise of the panel members and the research groups.

**Table 7 Euclidean distances between SAPV of Biomedical Sciences individual research groups, panel members, panel and groups together in the journal similarity matrix.**

|      | Groups | BIOM-A | BIOM-B | BIOM-C | BIOM-D | BIOM-E | BIOM-F | BIOM-G | BIOM-H | BIOM-I | BIOM-J | BIOM-K | BIOM-L | BIOM-M | BIOM-N | BIOM-O |
|------|--------|--------|--------|--------|--------|--------|--------|--------|--------|--------|--------|--------|--------|--------|--------|--------|
| Panel | 0.004 | 0.004 | 0.004 | 0.004 | 0.006 | 0.005 | 0.003 | 0.009 | 0.008 | 0.003 | 0.008 | 0.005 | 0.009 | 0.006 | 0.010 | 0.007 |
| PM1  | 0.006 | 0.007 | 0.006 | 0.006 | 0.007 | 0.007 | **0.003** | 0.011 | **0.009** | **0.002** | 0.009 | **0.007** | **0.011** | 0.008 | **0.012** | **0.009** |
| PM2  | 0.005 | **0.004** | 0.006 | 0.007 | 0.008 | 0.007 | **0.003** | **0.008** | 0.010 | 0.005 | **0.005** | 0.006 | 0.009 | 0.006 | 0.010 | 0.007 |
| PM3  | 0.007 | 0.007 | 0.006 | 0.007 | 0.008 | 0.008 | 0.006 | 0.011 | 0.011 | 0.006 | **0.008** | 0.006 | **0.011** | 0.008 | **0.011** | **0.009** |
| PM4  | 0.007 | 0.007 | 0.007 | 0.007 | 0.007 | 0.008 | **0.004** | 0.011 | 0.010 | **0.002** | 0.009 | 0.007 | **0.011** | 0.009 | **0.012** | **0.009** |
| PM5  | 0.004 | **0.005** | **0.002** | **0.003** | 0.007 | 0.006 | 0.005 | **0.009** | 0.009 | 0.005 | 0.008 | **0.004** | **0.009** | **0.006** | **0.009** | **0.006** |
| PM6  | 0.006 | 0.008 | 0.008 | 0.007 | 0.008 | **0.003** | 0.009 | **0.009** | **0.006** | 0.009 | 0.012 | 0.008 | **0.011** | 0.009 | **0.011** | **0.009** |
| PM7  | 0.007 | 0.008 | 0.008 | 0.007 | **0.005** | 0.007 | 0.009 | **0.010** | 0.007 | 0.008 | 0.012 | 0.008 | **0.011** | 0.009 | **0.011** | 0.009 |
| PM8  | 0.011 | 0.009 | 0.012 | 0.013 | 0.014 | 0.013 | 0.013 | **0.009** | 0.014 | 0.014 | 0.011 | 0.011 | **0.010** | 0.010 | **0.010** | 0.010 |

Shortest distances between a group and a panel member are underlined and printed in bold. Average shortest distances is 0.005 (SD 0.002). Distances whose confidence intervals overlap with that of shortest distance are in bold.

*Veterinary Science*

The Veterinary panel is the closest to VETE-B (0.005). The average shortest distances between the panel and individual research groups is 0.005 (SD 0.002). In the Veterinary department, the panel members are quite close to the research groups except for PM3 and PM4 (Table 8). PM3 and PM4 could be replaced with other potential panel members who have publications in journals that are more closely related to the groups' publication profile to obtain a better panel fit.

**Table 8 Euclidean distances between SAPV of Veterinary Sciences individual research groups, panel members, panel and groups together in the journal similarity matrix.**

|       | Groups | VETE-A | VETE-B | VETE-C |
|-------|--------|--------|--------|--------|
| Panel | 0.007  | 0.008  | 0.005  | 0.006  |
| PM1   | 0.011  | 0.013  | 0.010  | **0.005** |
| PM2   | 0.005  | **0.005** | **0.005** | 0.011 |
| PM3   | 0.015  | 0.016  | 0.013  | 0.013  |
| PM4   | 0.010  | 0.010  | 0.010  | 0.015  |

Shortest distances between a group and a panel member are underlined and printed in bold. Average shortest distances is 0.005 (SD 0.000). There are no shortest distances whose confidence intervals overlap with the other distances.

*Pharmaceutical Sciences*

Table 9 provides data on the distances between the Pharmaceutical Sciences panel and individual research groups. The average shortest distances between the panel and individual research groups is 0.008 (SD 0.042). PHAR-E (0.004) and PHAR-H (0.005) are closer to the panel while PHAR-I (0.013) is located moderately far away from all panel members except PM2. PHAR-I (0.011) and the panel do not share any common journals, but PM2 is also closer to this group than other panel members.

**Table 9 Euclidean distances between SAPV of Pharmaceutical Sciences individual research groups, panel members, panel and groups together in the journal similarity matrix.**

|       | Groups | PHAR-A | PHAR-B | PHAR-C | PHAR-D | PHAR-E | PHAR-F | PHAR-G | PHAR-H | PHAR-I | PHAR-J |
|-------|--------|--------|--------|--------|--------|--------|--------|--------|--------|--------|--------|
| Panel | 0.003  | 0.009  | 0.008  | 0.007  | 0.009  | 0.004  | 0.007  | 0.017  | 0.005  | 0.013  | 0.011  |
| PM1   | 0.013  | **0.011** | **0.011** | 0.017 | 0.015 | 0.015 | **0.008** | **0.020** | 0.012 | 0.021 | 0.020 |
| PM2   | 0.005  | 0.012  | **0.010** | <u>**0.005**</u> | 0.011 | <u>**0.004**</u> | 0.011 | 0.018 | 0.008 | <u>**0.011**</u> | <u>**0.008**</u> |
| PM3   | 0.006  | **0.010** | **0.009** | 0.009 | <u>**0.007**</u> | 0.007 | **0.008** | 0.018 | 0.007 | 0.015 | **0.013** |
| PM4   | 0.006  | **0.010** | **0.008** | 0.009 | 0.011 | **0.007** | **0.006** | 0.018 | <u>**0.007**</u> | 0.014 | **0.012** |
| PM5   | 0.007  | <u>**0.007**</u> | **0.008** | 0.010 | 0.012 | 0.008 | **0.007** | <u>**0.017**</u> | 0.007 | 0.017 | **0.014** |

Shortest distances between a group and a panel member are underlined and printed in bold. Average shortest distances is 0.008 (SD 0.004). Distances whose confidence intervals overlap with that of shortest distance are in bold.

*Biology*

The Biology panel is closer to BIOL-A (0.005) and BIOL-G (0.006), while BIOL-B (0.010) and BIOL-C (0.012) are at least 2 times farther away from the panel (Table 10). The average of the shortest distances between the Biology groups and panel members is 0.006.

**Table 10 Euclidean distances between SAPV of Biology individual research groups, panel members, panel and groups together in the journal similarity matrix.**

|       | Groups | BIOL-A | BIOL-B | BIOL-C | BIOL D | BIOL-E | BIOL-F | BIOL-G | BIOL-H | BIOL-I |
|-------|--------|--------|--------|--------|--------|--------|--------|--------|--------|--------|
| Panel | 0.004  | 0.005  | 0.010  | 0.012  | 0.007  | 0.008  | 0.007  | 0.006  | 0.010  | 0.008  |
| PM1   | 0.004  | 0.007  | **0.009** | 0.013 | <u>**0.004**</u> | 0.008 | 0.008 | **0.006** | <u>**0.009**</u> | 0.008 |
| PM2   | 0.005  | <u>**0.003**</u> | 0.010 | 0.015 | 0.009 | <u>**0.005**</u> | 0.012 | **0.005** | 0.011 | 0.012 |
| PM3   | 0.009  | 0.011  | 0.015  | **0.013** | 0.011 | 0.014 | <u>**0.003**</u> | 0.013 | 0.015 | <u>**0.004**</u> |
| PM4   | 0.007  | 0.006  | **0.009** | 0.016 | 0.009 | **0.006** | 0.014 | <u>**0.004**</u> | 0.010 | 0.013 |
| PM5   | 0.009  | 0.009  | 0.013  | <u>**0.012**</u> | 0.012 | 0.012 | 0.008 | 0.011 | 0.014 | 0.009 |

Shortest distances between a group and a panel member are underlined and printed in bold. Average shortest distances is 0.006 (SD 0.003). Distances whose confidence intervals overlap with that of the shortest distance are in bold.

### 4.3 Confidence intervals

To get an idea of the reliability of our barycenter and SAPV distances, we apply a bootstrapping approach to obtain 95% confidence intervals (CIs). Comparison of the CIs can then inform the analysis. Specifically, if two distances are not equal but their CIs overlap, the difference may not be meaningful.

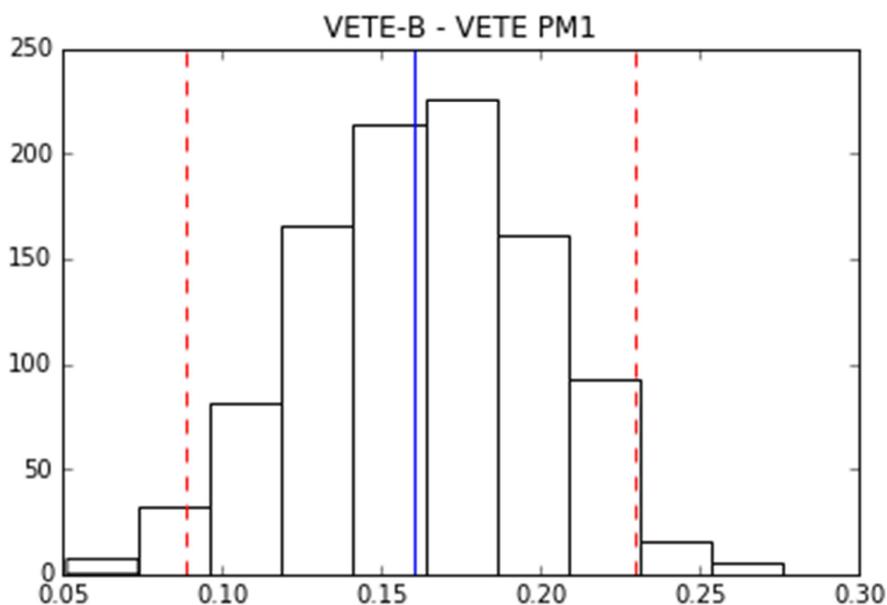

**Fig. 5 Histogram of 1000 bootstrapped distances between the barycenters of VETE-B and VETE-PM1 (Veterinary Sciences). The full line indicates the empirically found distance, the dashed lines indicate the CI.**

As explained in the Methods section, we calculate distances for 1000 bootstrap samples. The resulting distances tend to be normally distributed, as illustrated in Fig. 5. A similar image emerges for all disciplines and for both barycenters and SAPVs. It can be seen that the CI is a reliable approximation of the variability across the bootstrap samples.

We illustrate the interpretation of the CIs using a few examples. Our focus will be on the task of finding the panel members that are cognitively closest to a given research group. Fig. 6 displays the CIs for the distances between the barycenter of BIOM-D and the barycenters of all panel members in Biomedical Sciences. Ignoring the panel as a whole, the panel member for which we find the closest distance to BIOM-D is PM1 but we cannot simply conclude that this panel member is cognitively closest to the group: both PM4 and PM5 have CIs that partially overlap with PM1. Hence, PM4 and PM5 should be treated as viable alternatives to PM1 if one is seeking a panel member with expertise similar to that of research group BIOM-D.

Likewise, Fig. 7 displays CIs for SAPV distances, using the example of Pharmaceutical Sciences research group PHAR-A. In this case, it turns out that the differences between the panel members are relatively small. The result is that, with the exception of PM2, all panel members are eligible candidates. CI plots like Fig. 6 and Fig. 7 are available as supplementary online material for all research groups and for both barycenters and SAPVs.

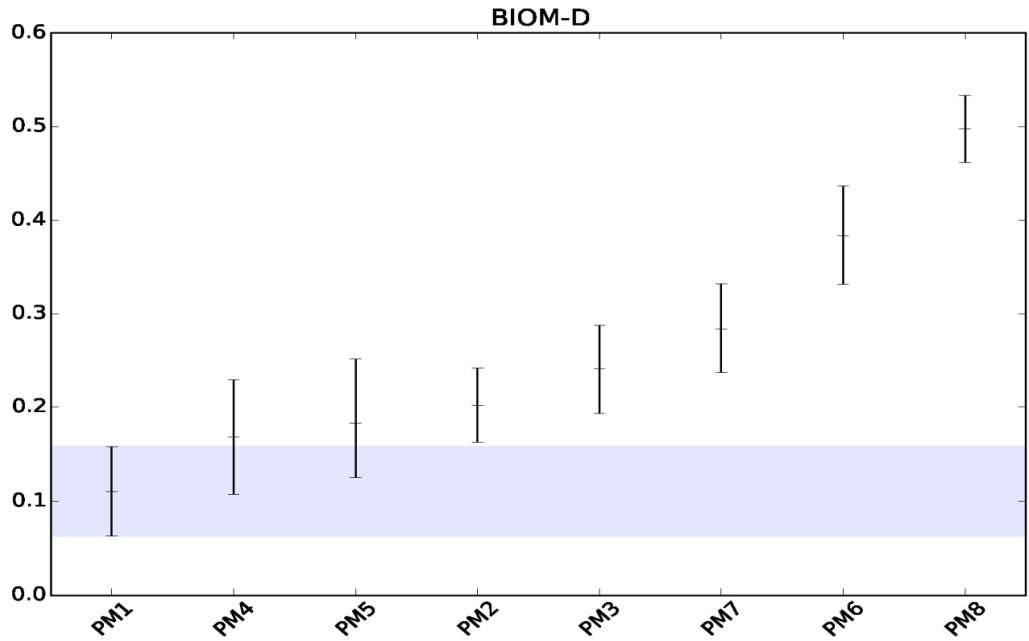

**Fig. 6** Confidence intervals for barycenter distances for Biomedical Sciences research group D. The highlighted part indicates the confidence interval of the shortest distance to the research group.

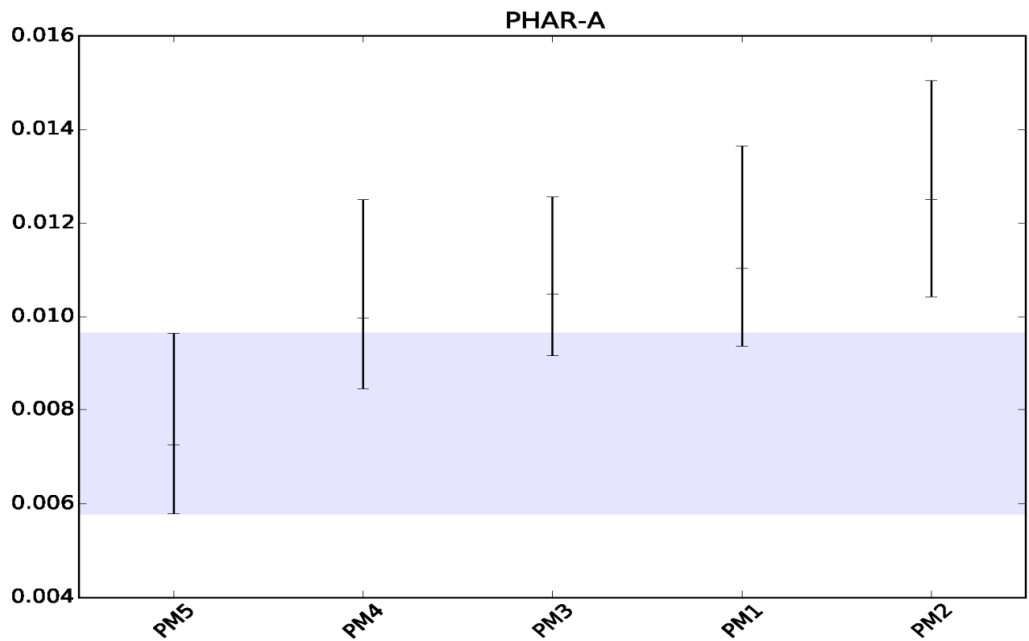

**Fig. 7** Confidence intervals for SAPV distances for Pharmaceutical Sciences research group A. The highlighted part indicates the confidence interval of the shortest distance to the research group.

We calculated the rate of overlap of CIs in the case of the barycenter approach and the case of the SAPV approach in all the four departments (see Table 11) in order to get a feel of the extent they might give rise to different conclusions. Overall, the degree of overlap due to the CIs of the barycenter approach seems similar to that of the SAPV approach.

**Table 11** Percentage of overlapping CI's for barycenters and SAPVs in each of the four disciplines.

| Department | Barycenter approach | SAPV approach |
|---|---|---|
| Biomedical Sciences | 36% | 34% |
| Veterinary Sciences | 44% | 0% |
| Pharmaceuticals Sciences | 43% | 55% |
| Biology | 28% | 28% |

*4.4 Comparison between two approaches*

To more directly compare the results we obtained from both approaches, we calculated the Pearson correlation coefficient (r) and the Spearman rank correlation coefficient (ρ) between the distances obtained through the barycenter approach and SAPV approach. The correlation calculation is based on all distances between research groups and individual panel members. Correlations for the Biomedical department (r = 0.60, ρ = 0.56), Biology department (r = 0.73, ρ = 0.71), Pharmaceutical department (r = 0.63, ρ = 0.62) and Veterinary department (r = 0.64, ρ = 0.66) are moderately strong (Fig. 8).

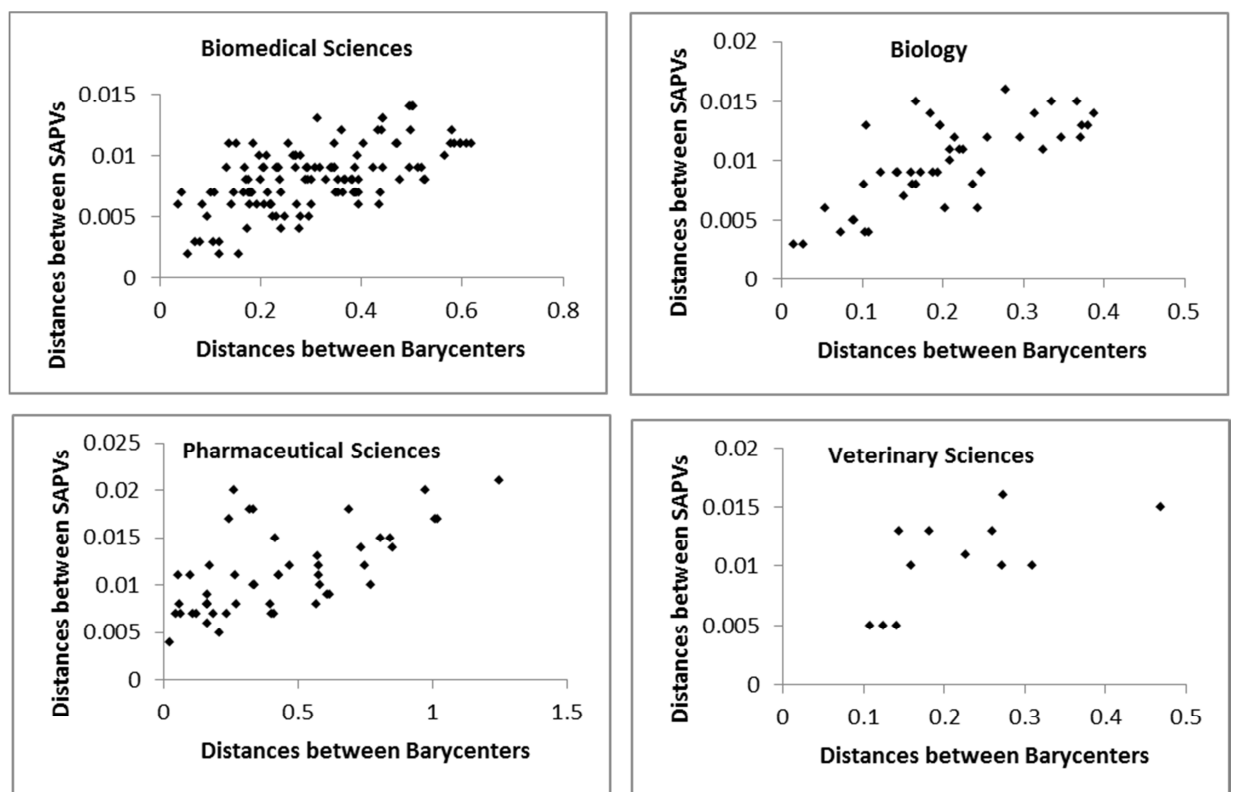

**Fig. 8** Scatter plot of the barycenter and SAPV distances between groups and individual panel members in the Biomedical Sciences, Biology, Pharmaceutical Sciences, and Veterinary Sciences departments.

We now turn to the question how the barycenter approach and the SAPV approach compare. Both try to quantify the cognitive distance by determining the Euclidean distance between representations or 'profiles' of an entity, but the way these profiles are obtained is quite different. The barycenter approach has the benefit of visualization, but the reduction of dimensionality that is inherent to creating a two-dimensional map may cause distortions in some cases. In this respect, the SAPV distances are the most reliable measure. We hypothesize that this advantage plays a larger role at the journal level than it did at the level of WoS categories, since there are many more dimensions in the former case. In general, we recommend using the SAPV approach for distance calculation and consider the barycenter approach more appropriate for visual exploration.

From the discussion on the composition of the four expert panels, it follows that a group can be far away from the panel as a whole. However, some individual panel members may have sufficient expertise to evaluate a single group, as indicated by publications in closely related or similar journals. For example, as discussed in section 4.1 and shown in Fig. 1, the barycenter of PM8 for Biomedical Sciences is in the immediate neighborhood of research groups BIOM-A, BIOM-J, BIOM-K and BIOM-L, while other panel members are farther away from them. On the other hand, according to the SAPV approach, BIOM-PM8 is situated moderately far away from all the research groups. In the same way, the barycenter of VETE-PM4 is far away from all the groups, while in the SAPV approach this is the case for PM3. These examples illustrate that, while the two approaches are clearly correlated, they may yield rather different results at the level of individual groups or panel members.

Even if a research group has no publications in the journals where the panel has publications, the panel might be able to evaluate the research group. For example, as discussed in section 4.1 and 4.2, there is no overlap between the journal portfolio of group PHAR-I and the Pharmaceutical Sciences panel, but PM2 is still fairly close to this research group (Fig. 3) both in the barycenter and the SAPV approach (Table 5 and Table 9).

Both the approaches give the opportunity to see how well fit the composition of the panel is if one or more panel members are replaced and compare the relative contribution of each potential panel member to the panel fit as a whole, by observing the changes to the distance between the panel's and the groups'. In future research, we intend to compare these approaches, as well as some others, with external data to gain more insight in their 'practical' merits.

## 5 Conclusion

We have considered two potential approaches of determining the match between research groups and expert panel members based on the journals in which they have published: distances (including confidence intervals) between barycenters on the map and distances (also including confidence intervals) between SAPVs. Both the barycenter and SAPV approaches hold serious advantages over a simple comparison of publication portfolios. Visualizations in the form of overlay maps can provide an intuitive picture of an entity's publication profile and include information on journal similarity, but they are less suited for actually distinguishing between, say, a few different panel members. In these cases, we have argued, distances between profiles that take similarity into account (like barycenters and SAPVs) constitute an approach with more 'actionable' results.

## 5.1 Discussion

A research group may deliberately hire other professionals, e.g., a biology research group might hire a physicist or computer scientist who continues to publish in their own discipline. In that case, the group's publication profile may change somewhat. We argue that it is the choice of the research group whether or not to include such publications in their research group profile during the period of research evaluation. As the formation of expert panel considers the focus of the research groups, the application of the barycenter and SAPV approaches are not affected.

In our case, the panel members have no prior involvement with the research groups, but the barycenter approach and SAPV approach can also be applied if the panel members have already collaborated with a research group or unit of assessment. The involvement of the panel member with the research group may result in a much better panel fit, but the research assessment itself might be subject to bias. However, such influence is outside the scope of our paper, as the formulation of criteria for selection of the panel members depends on the objectives of the concerned authority.

One might ask what distance between panel and research groups is acceptable for evaluation purposes. It seems to us that there is no *a priori* answer to this question, as the context, objectives and practical setting of an expert panel evaluation may all play a role. Hence, this cannot be decided beforehand. However, 'the shorter the distances the better the fit of the expert panel' can be suggested as a rule of thumb. At this point, we cannot make any claim regarding acceptable or preferable distances, and hence certainly not about the link between distances and the 'quality' of evaluations. In future research, we intend to address this issue, without, however, expecting to be able to set a norm.

Our proposed approaches help to identify expert panel members who have closely related expertise on the topic of the research group. Both approaches start from the publication profile of both the panel members and the research groups, assuming that these publication profiles adequately represent what they do. Therefore our proposed approaches might be less acceptable in some fields, e.g. the Engineering sciences, computer science, or social science and humanities, where non-journal outputs represent a larger part of the total output (see e.g. Rahm (2008) on computer science and Engels, Ossenblok, & Spruyt (2012) on SSH).

The scope of journals can vary significantly; some journals focus on rather specific topics, whereas others, such as PLoS ONE, are multidisciplinary in nature. One might therefore question whether journals are the adequate level of analysis. We suggest two possible routes for future research in this regard. First, it would be interesting if a comparison could be made between an analysis that considers all journals and one that leaves out multidisciplinary or otherwise broadly scoped journals. Second, one could replace journals with clusters of cognitively related articles. For instance, one could use the CWTS (Centre for Science and Technology Studies) article-level classification (Waltman & van Eck, 2012), which groups related articles together on the basis of direct citations regardless of the journal in which they were published. While we consider this an interesting idea, we also point out that it harbors its own set of theoretical and practical problems.

*5.2  Normative implications*

Our proposed expert panel composition methods based on journal data allow the panel composition authority to see in advance about the panel's fit to the research groups that are going to be evaluated. The distance between units of assessment can be used as an indicator of cognitive distance. Therefore, the concerned authority will have the opportunity to replace outliers among the panel members to make the panel fit well with the research groups to be evaluated. For example, the authority can find a best-fitting expert panel by replacing a more distant panel member with a potential panel member located closer to the groups, in addition to the other panel member to cover the expertise of the PHAR-I research group. Also, the distances between panel members and research groups could be used to facilitate the division of labor among the panel members. In our opinion, adequate coverage can be considered a necessary condition for the quality of an evaluation.

Both the barycenter and SAPV approaches to measuring cognitive distance can be used to inform the process of expert panel composition for a collection of research groups. Rahman et al. (2015) applied the barycenter approach on a global map of science based on Web of Science subject categories. In this study at the level of journals, we have applied both barycenter and SAPV approaches. Our future research will focus on the difference between these two approaches in WoS subject categories and journals, and lead us to a comprehensive approach to expert panel composition.

**Acknowledgments**

The authors thank Ronald Rousseau for stimulating and insightful suggestions related to the topic of the paper and Thomson Reuters for making the journal citation data available. This investigation has been made possible by the financial support of the Flemish government to ECOOM, among others. The opinions in the paper are the authors' and not necessarily those of the government. We thank the reviewers for their constructive remarks.